\documentclass[11pt]{article}

\usepackage[utf8]{inputenc}
\usepackage[T1]{fontenc}
\usepackage{lmodern}
\usepackage{upgreek}
\usepackage{bm}
\usepackage{amsmath, amssymb, amsfonts}
\usepackage{graphicx}
\usepackage{hyperref}
\usepackage{geometry}
\usepackage{tabularx}
\usepackage{physics}
\usepackage{siunitx}
\usepackage{cite}

\title{Focusing and light collection effects on plasma-induced frequency-resolved optical switching (PI-FROSt) traces}

\author{
Mario Guerras, Íñigo J. Sola, Benjamín Alonso, Enrique Conejero Jarque,\\
Ignacio Lopez-Quintas, Julio San Roman, and Aurora Crego\\
{\footnotesize Grupo de Investigación en Aplicaciones del Láser y Fotónica, Física Aplicada, Universidad de Salamanca, Spain} \\
{\footnotesize Unidad de Excelencia en Luz y Materia Estructuradas (LUMES),
Universidad de Salamanca, Spain} \\
}

\begin{document}

\maketitle

\begin{abstract}
Plasma-Induced Frequency-Resolved Optical Switching (PI-FROSt) is a promising and recently proposed phase-matching-free technique for characterising ultrafast pulses across broad spectral ranges. We investigate the mechanisms of PI-FROSt trace formation through numerical simulations and experimental validation. The results reveal that trace characteristics are highly sensitive to the relative focusing geometry between pump and probe pulses, as well as the spatial region selected for signal collection. Depending on these conditions, the interplay between plasma defocusing and positive lens-like nonlinear effects causes either intensity depletion or enhancement in the probe beam, flipping the PI-FROSt trace. Simulations demonstrate that optimal gate stability also depends strongly on the focusing scheme and the collecting region. This study highlights that precise spatial and temporal optimisation is essential to properly exploit the benefits of this broadband pulse characterisation technique.
\end{abstract}


\section{Introduction}
Ultrafast pulses in the femtosecond domain play an important role in a number of diverse fields, such as ultrafast spectroscopy, attoscience, microscopy, and materials processing, among others. The capability to measure the shape and phase of these pulses is essential for understanding and controlling light-matter interactions on ultrafast timescales. In the last decades, several techniques for pulse characterisation have arisen and are widely used, e.g., FROG (Frequency-Resolved Optical Gating) \cite{Kane1993}, SPIDER (Spectral Phase Interferometry for Direct Electric-field Reconstruction) \cite{Iaconis1998}, MIIPS \cite{Lozovoy2004}, d-scan \cite{Miranda2012}, a-swing \cite{Alonso2020}, etc. Most of them are based on nonlinear signal generation through parametric optical processes (with second harmonic generation being the most commonly used effect, although other nonlinear processes such as third harmonic generation can be also used). Common challenges in these processes include ensuring a proper phase matching to avoid distortions and artefacts in the measurement, as well as detecting the generated signal for the characterisation process, since it could lie in the ultraviolet region when working with pulses in the visible range. In addition, phase matching can be particularly critical in broadband laser pulses, and it is not always easily achievable.

Frequency-Resolved Optical Switching (FROSt) \cite{Leblanc2019,Lassonde2021} is a technique that focuses on overcoming this limitation. In FROSt, as in the FROG technique, of which it could be considered a particular variant, the gating concept is employed: a pump pulse is used to excite free carriers in a semiconductor sample, creating a sharp absorption edge in time that acts as an optical gate to characterise a probe pulse. The spectrum of the probe pulse is measured as a function of the delay between the pump and the probe, allowing the electric field of the probe pulse to be reconstructed, for example using a ptychographic algorithm. FROSt offers several advantages over other pulse characterisation techniques, such as independence from phase-matching constraints \cite{Leblanc2021}, high phase change sensitivity \cite{Longa2022}, the ability to characterise test pulses with very low energy (as long as a synchronized powerful pump beam is available) and arbitrarily long duration, while adapting to high repetition rate scenarios \cite{Haddad2023}. This behaviour makes the technique successful in the infrared and low frequency visible regions. However, this technique is limited in its spectral range by the transparency of the material used for the pulse under test. The lack of transparent materials in some parts of the visible region and the ultraviolet spectral region restricts the ability of this technique to characterize such pulses. These limitations have motivated the continuous search for alternative approaches that balance versatility and simplicity.

One way to overcome the transparency limitations of semiconductors is to use a different phenomenon to create sharp optical switching. One alternative consists of using the rise of plasma density induced by ultrashort pulses in gasses. In a first approach, called Plasma Mirror FROG (PM-FROG) \cite{Itakura2015}, the plasma mirror effect (appearing when an intense ultrafast laser pulse interacts with a solid surface, ionizing it to create a plasma that reflects subsequent pulses and acts as a rapidly forming optical mirror) can be integrated into the FROG scheme. Thus, the plasma mirror is formed by irradiating a fused silica surface with an intense laser pulse. This transient mirror acts as an ultrafast optical switch and reflects the probe pulse only when the delay between the pump and the probe is such that the plasma mirror is present. The spectrum of the reflected probe pulse is measured as a function of the delay, and the probe pulse waveform can be recovered from the resulting spectrogram using a FROG algorithm. PM-FROG has been used to characterise vacuum ultraviolet pulses with durations as short as 20 fs \cite{Itakura2019}. However, PM-FROG requires a complex setup and alignment procedure.

More recently, a new technique called Plasma-Induced FROSt (PI-FROSt) has been developed \cite{Bhalavi2024,bejot2024}. In PI-FROSt, a plasma lens is generated through a non-resonant multiphoton ionisation process by focusing a moderately intense pump pulse into a gas, typically argon, or even ambient air. The ultrafast build-up of the plasma lens allows it to operate as a switch in the femtosecond time scale, diffracting part of a probe pulse that is focused in a manner similar to the pump. The modified portion of the emerging probe beam is isolated in the far field using a coronagraph and measured with a spectrometer as a function of the delay between the pump and probe pulses. This produces a trace that enables full characterisation of the probe’s temporal and spectral features. PI-FROSt is a simple, phase-matching-free technique that can operate across the gas transparency range, without requiring frequency doubling or tripling the probe spectrum. It has been used to characterise pulses in both the near-infrared and ultraviolet spectral ranges \cite{Bejotpreprint}.

The objective of the present work is to study from a fundamental point of view the PI-FROSt technique to obtain further insight into its potential and limitations. For this purpose, in Section 2, we present the experimental setup. Secondly, the developed (2+1)D numerical model of plasma build-up and its effect on the probe pulse, including the main features that describe the propagation and interaction of ultrashort pulses in air, is presented in Section 3. Once both the model and experimental setup have been introduced, the technique is studied under diverse experimental and numerical conditions (Section 4) to establish the physical fundamentals of the technique. Furthermore, the impact of several critical parameters on technique performance will be studied. Along this analysis, two types of signal collection of the probe beam are considered: at the periphery, as previously reported in the literature, and at the centre of the probe beam. Finally, in Section 5, we summarise the main results of the work and their implications for future research.

\section{Experimental Layout}
The laser source for the experiment was a commercial CPA-amplified Ti:sapphire laser chain (Spitfire ACE from Spectra Physics). This setup delivered pulses with a central wavelength of 800 nm at a repetition rate of 5 kHz. The pulses, linearly polarised, had a duration of 60 fs, close to the Fourier limit, and an energy of up to 1.6 mJ with a beam waist of 5 mm. 

\begin{figure}[htbp]
\centering\includegraphics[width=10cm]{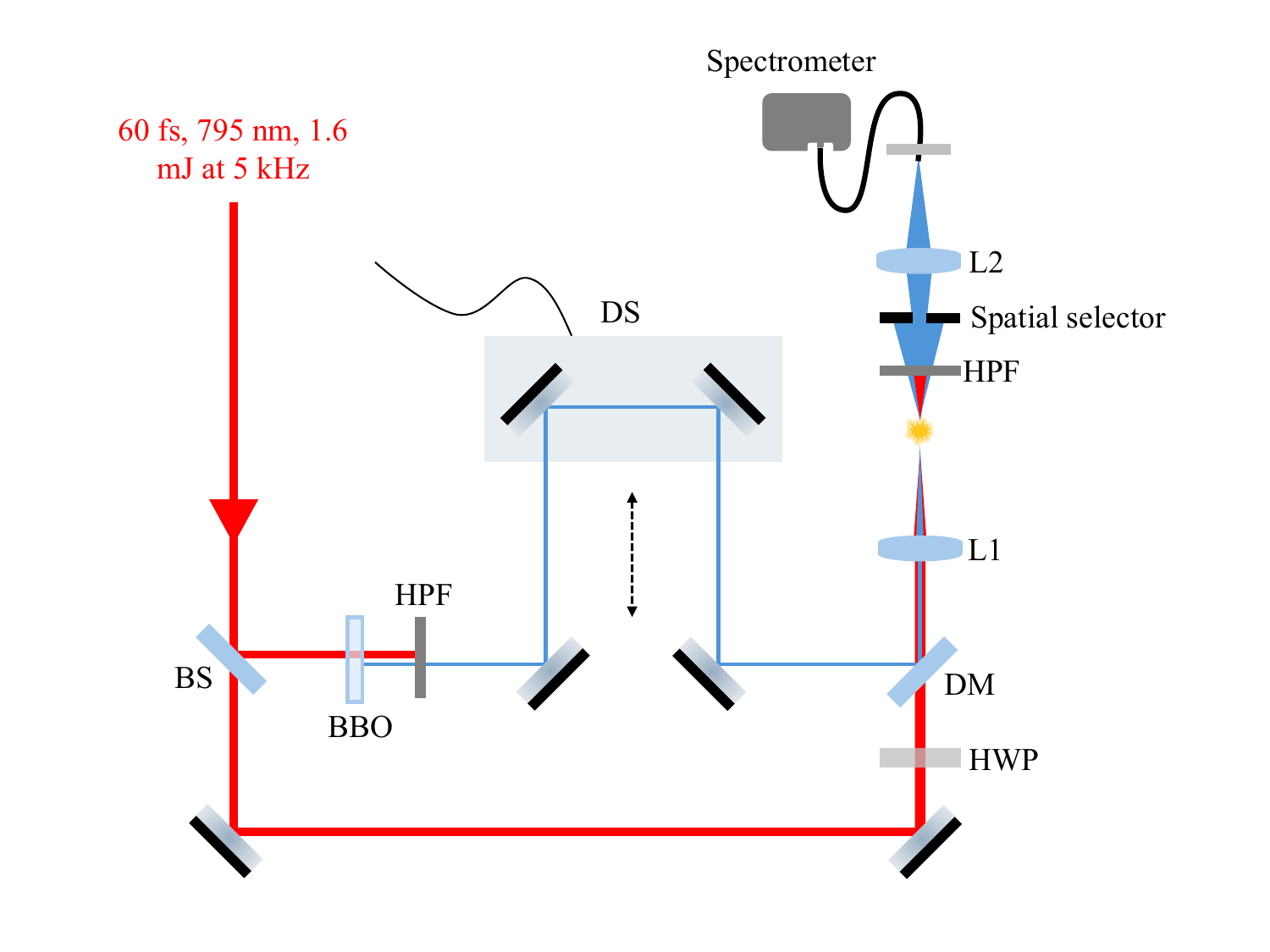}
\caption{Scheme of the PI-FROSt setup. The output of a CPA-amplified Ti:sapphire laser is split into two arms by a beam splitter (BS). The transmitted beam serves as the pump pulse, with a half-wave plate (HWP) controlling its polarisation orientation. The reflected beam is frequency-doubled in a BBO crystal, and a high-pass filter (HPF) removes the residual fundamental beam. Both beams are recombined using a dielectric mirror (DM), while a delay stage (DS) in the probe path adjusts the temporal overlap. Finally, a focusing element (L1), lens or mirror depending on the experimental conditions, brings both pulses to a focus and the signal is collected into a spectrometer by a collecting lens (L2). Prior to L2, the pump is filtered (HPF) and an element for the spatial selection of the collected light at the central or peripheral part of the beam is placed (spatial selector).}
\label{Fig1}
\end{figure}

The PI-FROSt pump-probe setup consists of an interferometer (Fig. \ref{Fig1}) in which the incoming laser beam (\(140\,\upmu\mathrm{J}\) per pulse) is split into two arms by a 50/50 beam splitter (BS). The transmitted beam acts as the pump pulse to generate the plasma gate. An achromatic half-wave plate (HWP) in this arm controls the linear polarisation orientation of light to perform the experiment with the pump and probe polarisations oriented either parallel or perpendicular to each other. To minimise strong XPM effects, the experiments are performed using perpendicular pump and probe polarisation. The probe pulse is generated by frequency doubling the incident pulse in a BBO crystal, and the residual fundamental frequency signal is removed with a high-pass filter (HPF). A delay stage (DS) is incorporated into this arm to control the relative delay between the pump and probe pulses. The pump and probe beams, with pulse energies of \(E_{pump}=82\,\upmu\mathrm{J}\) and \(E_{probe}=0.4\,\upmu\mathrm{J}\), are then collinearly recombined using a dichroic mirror (DM) and pass through a focusing element (L1). The plasma gate is created at the focus position of L1 by the pump pulse and affects the probe beam depending on their relative delay.

We experimentally explore the impact of the relative position of the pump focus (plasma position) and probe focus on the PI-FROSt trace build-up. We consider two focusing configurations depending on the element used as L1: a singlet lens (LA4021 from Thorlabs), which presents chromatic aberration; thus, the probe exhibits a shorter focal length at $\lambda = 400\,\mathrm{nm}$ ($f(400)\approx 29.60\,\mathrm{mm}$) than the pump at $\lambda = 800\,\mathrm{nm}$ ($f(800)\approx 30.34\,\mathrm{mm}$), according to the lens specifications and the Lens Maker's Formula; or a mirror, a spherical Ag mirror ($f$ = 30 mm), which has the same focal length for both pump and probe. Short focal lengths are employed to enhance the effect, as they result in higher peak intensities at focus.

Finally, the pump pulse is filtered again (HPF), and the resulting probe pulse radiation is collected by means of a collecting lens (L2), coupling the light into an optical fibre and a spectrometer (Ocean Optics HR4000). Depending on the signal collection configuration, as will be discussed later, an iris or a central blocker is used just before L2 to select the centre or the peripheral part of the probe beam (spatial selector). The PI-FROSt traces are obtained by scanning the relative delay between the probe and pump beams. This is achieved by placing the delay stage at different positions and measuring the probe pulse beam at each delay step. Before doing all the measurements, we verified that the pump did not generate any relevant second harmonic signal in air by itself.

\section{Numerical model}
In this section, we present the numerical model that describes the complete spatiotemporal dynamics of both pump and probe pulse during propagation, from the nonlinear wave equation for the following different optical components \cite{agrawal2001}
\begin{equation}
\bm{E}(\bm{r},t,z)
=
\bm{E}_1(\bm{r},t,z)
+
\bm{E}_2(\bm{r},t,z)
=
\bm{A_1}(\bm{r},t,z)e^{i(k_1 z-\omega_1 t)}
+
\bm{A_2}(\bm{r},t,z)e^{i(k_2 z-\omega_2 t)}
\end{equation}
where \( k_j = \frac{n_L(\omega_j)\,\omega_j}{c} \) for \( j=1,2 \), is the wave number, \( \omega_j \) is the central frequency, and \( n_L(\omega_j) \) is the value of the linear refractive index for \(\omega_j \), \( c \) is the speed of light in vacuum, and \( \bm{r} \),\( z \) and \( t \) denote the transverse, longitudinal (propagation) and temporal coordinates, respectively. 

We model the propagation of two pulses in a pump-probe configuration, where the pump pulse, \(\bm{E}_1(\bm{r},t,z) \), has enough intensity to propagate nonlinearly through the medium, ionise it and generate a plasma. The probe pulse, \(\bm{E}_2(\bm{r},t,z) \), is a sufficiently attenuated pulse that propagates linearly and interacts nonlinearly with the pump pulse through the generated plasma and the cross-phase modulation (XPM) contributions from the electronic and molecular response. We assume that both beams are linearly polarised with orthogonal polarisations, the cylindrical symmetry is preserved, and the scalar envelopes, \(\bm{A_1} (\bm{r},t,z) = A_1 (r,t,z)\bm{u_x} \equiv A_1x\) and \(\bm{A_2} (\bm{r},t,z) = A_2 (r,t,z)\bm{u_y} \equiv A_2y\), slowly vary along the propagation distance, \( z \). Using this approach, we obtain a set of equations for the two envelopes to model their propagation dynamics that are coupled by the third-order nonlinear response of the medium and the ionisation. Assuming cylindrical symmetry and using a frame moving with the pump pulse, \( t' = t - k_{11} z \), where \( k_{1j} = v_{gj}^{-1} = \left(\frac{\partial k}{\partial \omega}\right)_{\omega_j} \), for \( j = 1,2 \), the propagation equations for the PI-FROSt scheme in air can be expressed as \cite{COUAIRON2007, boyd2020}
\begin{align}
\Bigg(
-i\frac{\nabla_\perp^2 \hat{T}_1^{-1}}{2k_1}
+ \frac{\partial}{\partial z}
- iD_1
+ \frac{\alpha_1}{2}
\Bigg) A_{1x}
&= i\gamma_1 \hat{T}_1 \Bigg(
(1-f_R)|A_{1x}|^2 + \nonumber \\
&\quad + \frac{f_R}{\tau_K}
\int_{-\infty}^{t'}
e^{-\frac{t'-\tau}{\tau_K}}
|A_{1x}(\tau)|^2\, d\tau
\Bigg) A_{1x} - \nonumber \\
&\quad - \frac{\sigma_1}{2}(i\omega_1\tau_C+1)\hat{T}_1^{-1} (\rho A_{1x})
 - \frac{W(|A_{1x}|^2)U_i}{2|A_{1x}|^2}(\rho_{at}-\rho)
 \label{Eq2}
\end{align}

\begin{align}
\Bigg(
-i\frac{\nabla_\perp^2 \hat{T}_2^{-1}}{2k_2}
+ \frac{\partial}{\partial z}
- iD_2
+ \frac{\alpha_2}{2} + (k_{12}-k_{11})\frac{\partial}{\partial t'}
\Bigg) A_{2y}
&= i\gamma_2 \hat{T}_2f_{XPM}\Bigg(
(1-f_R)2|A_{1x}|^2 +\nonumber \\
&\quad + \frac{f_R}{\tau_K}
\int_{-\infty}^{t'}
e^{-\frac{t'-\tau}{\tau_K}}
|A_{1x}(\tau)|^2\, d\tau
\Bigg) A_{2y} - \nonumber \\
&\quad - \frac{\sigma_2}{2}(i\omega_2\tau_C+1)\hat{T}_2^{-1} (\rho A_{2y})
\label{Eq3}
\end{align}
where \(\gamma_j = \frac{\omega_j}{c}\,\frac{n_L(\omega)\,n_{NL}}{n_L(\omega_j)}\) is the nonlinear coefficient, being \( n_L \) the linear refractive index and \( n_{NL} \) the nonlinear refractive index. The left-hand side of equations (\ref{Eq2})-(\ref{Eq3}) represents the linear propagation of both pulses through the medium, while the right-hand side describes the nonlinear effects occurring during the propagation of each beam. The first linear term accounts for both the diffraction of the beam in the transverse plane, represented by the transverse Laplacian \(\nabla_\perp^2 = \frac{1}{r}\frac{\partial}{\partial r} + \frac{\partial^2}{\partial r^2}\) and the spatiotemporal coupling described by the operator \(\hat{T}_j = \left[1 + \frac{i}{\omega_j}\frac{\partial}{\partial t'}\right]\) for a more accurate description of ultrashort pulse propagation \cite{COUAIRON2007, brabec1997}. The second term, \(D_j\), is the dispersion operator, \(D_j(t') = \sum_{m \ge 2}^{\infty} \frac{1}{m!}\, k_{jm} \left(i\frac{\partial}{\partial t'}\right)^m\), which includes all higher-order dispersion terms, being \(k_{jm} = \left(\frac{\partial k}{\partial \omega}\right)^m_{\omega_j}\), and the third term related to \(\alpha_j\) accounts for the linear absorption suffered by each beam when propagating through the medium. The last term on the left-hand side of equation (\ref{Eq3}) is associated with the group-velocity mismatch, since both beams are propagating in the frame of the pump pulse.

The nonlinear propagation varies in each case. In equation (\ref{Eq2}), the pump pulse is assumed to have a sufficiently high intensity to self-induce different nonlinear effects: self-focusing, self-phase modulation, Raman contributions from the molecular response function of the gas \cite{Couairon2002} and plasma generation. The complete Kerr response combines the instantaneous and delayed responses of the medium, where \(f_R\) denotes the relative weight of the Raman contribution and \(\tau_K\) its characteristic response time. The self-steepening effect is taken into account through the operator \(\hat{T}\) in the nonlinearity. The last terms on the right-hand side of Eq. (\ref{Eq2}) describe the plasma generation, which becomes significant as the pulse intensity reaches \(10^{13}-10^{14} W/cm^2\). Once ionisation occurs, the generated free electrons give rise to plasma absorption, defocusing, and blueshift of the pulse spectrum. These effects depend on the collision time \(\tau_C\) and on the inverse Bremsstrahlung cross section \(\sigma_j\), given by \(\sigma_j=(\omega_j/n_L(\omega)c\rho_c)\cdot\omega_j\tau_C/(1+\omega_j^2\tau_C^2)\), where \(\rho_c=\epsilon_0 m_e \omega_j^2/e^2\) is the critical plasma density \cite{COUAIRON2007}. The last term associated with the ionization of the medium represents the pump energy losses related to the plasma generation process. The ionisation rate \(W(|A_1 |^2 )\) depends on the ionisation potential \(U_i\), the laser intensity, and the central frequency of the pulse. Equations (\ref{Eq2})-(\ref{Eq3}) are solved simultaneously with an evolution equation (Eq. (\ref{Eq4})) for the electron density, \(\rho\). 
\begin{equation}
\frac{\partial \rho}{\partial t'}
=
W\!\left(|A_1|^2\right)\left(\rho_{\mathrm{at}}-\rho\right)
\label{Eq4}
\end{equation}
Here, \(\rho_{at}\) is the atomic density of the neutral gas (air in the present work). We account only for the ionisation of oxygen and neglect nitrogen, since oxygen has a lower ionisation potential and contributes more strongly to plasma dynamics at these intensities \cite{COUAIRON2007,Couairon2002}. This assumption remains valid as long as the plasma electron density stays below 20\% of the total density \(\rho_{at}\), which corresponds to the oxygen fraction in air \cite{COUAIRON2007}. Thus, in this model, Eq. (\ref{Eq4}) describes exclusively oxygen ionisation. We also neglect avalanche and recombination processes in electron-density evolution, keeping only multiphoton ionisation, which is a good approximation for femtosecond pulses \cite{Couairon2002}.

As mentioned earlier, each pulse experiences a different nonlinear response during propagation. Under the weak-probe approximation, Eq. (\ref{Eq3}) accounts only for the nonlinear effects induced by the pump pulse, which are the XPM effects (electronic and molecular responses) and the refractive index modifications induced by the plasma generated by the pump pulse. Through XPM, the pump pulse induces a change in the probe, with a strength quantified by the factor \(f_{XPM}\), which depends on the polarisation of both beams. For linearly polarised beams in an isotropic medium, \(f_{XPM}=1/3\) when they are orthogonally polarised \cite{agrawal2001,boyd2020}. 
We are interested in solutions of the coupled equations for input pump and probe pulses modelled as cylindrically symmetric Gaussian beams, where each beam is expressed as
\begin{equation}
A(r,t,0)
=
A_0
\exp\!\left(-\frac{r^2}{w_0^2}\right)
\exp\!\left(-\frac{t^2}{t_p^2}\right)
\exp\!\left(-\frac{i k r^2}{2f}\right)
\end{equation}

Here, \(A_0\) is the amplitude, \(w_0\) is the beam waist, \(t_p\) is the pulse duration and \(f\) is the focal length of the focusing element. To solve Eqs. (\ref{Eq2})-(\ref{Eq4}), we have developed a (2+1)D numerical model based on a split-step method \cite{agrawal2001}, which includes the spatial and temporal dynamics of the pulses during propagation, as described in detail in \cite{couairon2011,crego2023}. The linear part is computed at different steps: the dispersion is solved as an exponential operator in the frequency domain, while the diffraction and spatiotemporal coupling terms are solved using a Crank-Nicolson scheme applied to each spectral component. The nonlinear contributions and the evolution of the electron density are solved using a fourth-order Runge-Kutta algorithm in the spatiotemporal domain. At each propagation step, the pump-induced ionisation is calculated based on its local intensity and coupled back into Eqs. (\ref{Eq2})-(\ref{Eq3}) to solve the nonlinear dynamics. Ionisation rates are calculated using the PPT model, which accurately describes both the multiphoton and tunnel regimes \cite{perelomov1966}.

To be able to understand and interpret the experimental results, we carefully select the parameters for the simulations to reproduce the overall trend of the dynamics while keeping the computational cost manageable: an infrared (IR) unchirped pump pulse is centred at 800 nm and has a temporal duration of 60 fs (FWHM). The probe pulse, centred at 400 nm, is also assumed to exhibit a 60-fs duration (FWHM). In the present calculations, we consider a collinear focusing scheme. The simulation is initialized with a reduced beam size, effectively representing a pre-focused condition. Both beams have an initial beam radius of $w$ = 0.5 mm, which implies that, for the same focal length, the probe focal spot is half the size of the pump spot, ensuring that the entire probe beam interacts with the plasma. The pulse energies are adjusted so that the IR \( (E_{IR} = 50\,\upmu\mathrm{J}) \) induces plasma generation and XPM, while the SH remains at low energy \( (E_{SH} = 5\,\upmu\mathrm{J}) \) to avoid generating additional nonlinear effects. The equivalent focal length of the focusing element (L1) for the pump beam is set to \(f_{pump} = 300\,\mathrm{mm}\), while for the probe beam, two focusing scenarios are considered: \(f_{probe} = 270\,\mathrm{mm}\) and \(f_{probe} = 300\,\mathrm{mm}\).

The purpose of these simulations is to understand the underlying physics of the PI-FROSt technique and to interpret the features observed in the experimental traces. Additional simulation parameters are listed in Table \ref{tab1}. The expression for the linear refractive index of air is calculated using the formula from \cite{Zhang2008}. For the nonlinear refractive index, we assume that the dispersion is negligible, and the same value is used for both pulse wavelengths.
\smallskip

\begin{table}[htbp]
\caption{Parameters for modelling nonlinear pulse propagation in air at pump wavelength$^{a}$}
  \label{tab1}
  \centering
\begin{tabularx}{1\textwidth}{ >{\centering\arraybackslash}X  >{\centering\arraybackslash}X  >{\centering\arraybackslash}X  >{\centering\arraybackslash}X >{\centering\arraybackslash}X >{\centering\arraybackslash}X}
\hline
$p$ (bar) & $n_2$ (mm$^2$/W) & $f_R$ & $\tau_K$ (fs) & $\rho_{\mathrm{at}}$ (cm$^{-3}$) & $\tau_C$ (fs) \\
\hline
1 &
$3.2\times10^{-17}\,p$ &
0.5 &
70 &
$2.7\times10^{19}\,p$ &
350 \\
\hline
\end{tabularx}
$^a$ Values of \(n_2\), \(f_R\), \(\tau_k\) and \(\rho_{at}\) are obtained from \cite{Couairon2002}, $p$ denotes the air pressure, and \(\tau_C\) is obtained from \cite{kolesik2006}.
\end{table}

\section{Results and Discussion}
We combine numerical simulations and experiments to understand the physical phenomena underlying the PI-FROSt technique. This dual approach allows to reveal the nonlinear dynamics and interpret the experimental traces more accurately.

\subsection{Relevance of light collection region and relative focus position among plasma and probe.}
\begin{figure}[htbp]
\centering\includegraphics[width=15cm]{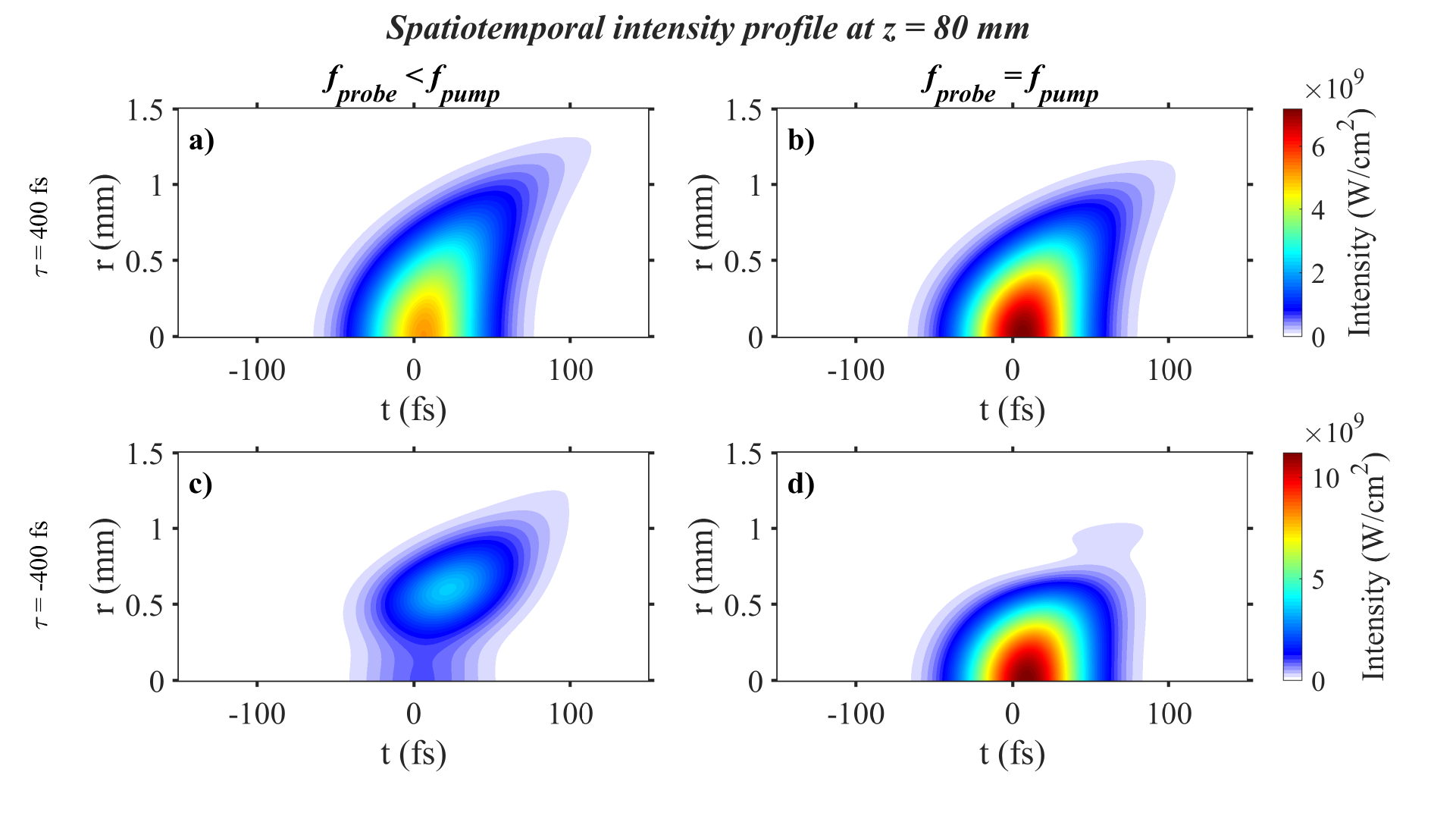}
\caption{Spatiotemporal intensity profiles of the probe pulse at $z = 80\,\mathrm{mm}$ after L1 for two different focusing configurations: $f_{probe} < f_{pump}$ (a and c), and $f_{probe} = f_{pump}$ (b and d). Each row shows the intensity profile for two pump-probe delays, 400 fs (a and b, probe arrives first) and -400 fs (c and d, pump arrives first). A single colorbar is shared across each row, representing the probe intensity scale for a given delay for a better comparison.}
\label{Fig2}
\end{figure}

To understand signal formation in PI-FROSt, we numerically investigated the influence of the pump, the plasma creation, and the focusing geometry on the probe spatiotemporal intensity profiles for different pump–probe time delays and the two focusing configurations commented above, which we denote for simplicity as $f_{probe} < f_{pump}$ and $f_{probe} = f_{pump}$. Figure \ref{Fig2} presents the evolution of the spatiotemporal intensity profile of the probe beam at 80 mm after the focusing element (L1), and 50 mm after pump focus, for these two focusing configurations: $f_{probe} < f_{pump}$ (Fig. \ref{Fig2}a and \ref{Fig2}c) and $f_{probe} = f_{pump}$ (Fig. \ref{Fig2}b and \ref{Fig2}d). Each row corresponds to a different pump-probe time delay: in the upper row (Fig. \ref{Fig2}a and \ref{Fig2}b), the probe arrives 400 fs before the pump, and its propagation remains essentially linear, while in the bottom row (Fig. \ref{Fig2}c and \ref{Fig2}d), the probe arrives 400 fs after the pump (and the plasma generated), which modifies the probe’s propagation differently depending on the relative focal positions.

In the focusing configuration where $f_{probe} < f_{pump}$ and the probe beam is travelling after the pump beam (Fig. \ref{Fig2}c), it experiences defocusing due to the effect of the plasma. Consequently, the central region of the probe beam (which we will refer to as “centre”) is depleted, while light is redistributed towards a more peripheral spatial zone (called “corona”). By contrast, in the focusing configuration where $f_{probe} = f_{pump}$, (Fig. \ref{Fig2}d), the effect on the subsequent probe propagation is different: the probe pulse, instead of being defocused by the generated plasma, experiences an increase in intensity along the propagation axis, leading to a light hotspot in the centre and depleting the corona. 

Throughout this work, we will refer to the depletion of light caused by the plasma on the probe beam in a given probe spatial region (centre or corona) as “shadow”, while the increase of light in a probe beam region will be called “surge”.

\begin{figure}[htbp]
\centering\includegraphics[width=15cm]{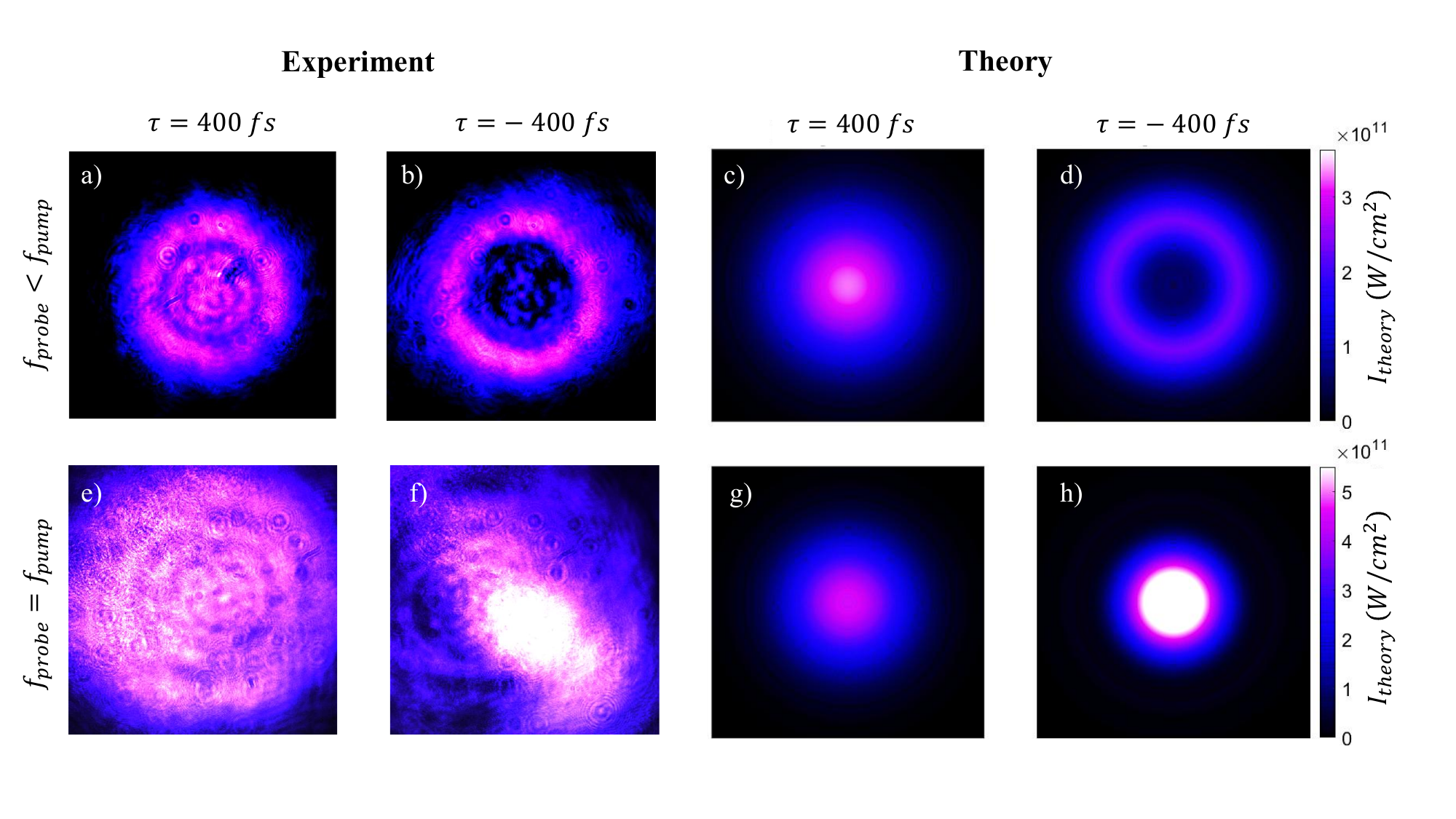}
\caption{Measured (a, b, e, f) and simulated (c, d, g, h) transverse intensity profiles of the probe, showing cases where the probe arrives before (a, c, e, and g, $\tau$ = 400 fs) or after (b, d, f, and h, $\tau$ = -400 fs) the pump pulse, for the two different probe focal length configuration: $f_{probe} < f_{pump}$ in panels a to d; and $f_{probe} = f_{pump}$ in panels e to h.}
\label{Fig3}
\end{figure}

To further illustrate this behaviour, Fig. \ref{Fig3} shows the experimental and simulated transverse intensity profiles of the probe beam (i.e., time integrated at $z = 80\,\mathrm{mm}$), under the mentioned spatio-delay conditions. As shown in Figure \ref{Fig2}, we again compare the two focusing configurations: $f_{probe} < f_{pump}$ (Fig. \ref{Fig3}a-\ref{Fig3}d); and $f_{probe} = f_{pump}$ (Fig. \ref{Fig3}e-\ref{Fig3}h). 

Both the experimental and simulated transverse intensity profiles of the probe beam shown in Fig. \ref{Fig3} are consistent with the dynamics previously observed in Fig. \ref{Fig2}. When $f_{probe} < f_{pump}$, the probe beam remains essentially unchanged when it arrives before the pump ($\tau$ = 400 fs, Fig. \ref{Fig3}a and \ref{Fig3}c), but a “central shadow” appears when it arrives after the pump ($\tau$ = -400 fs, Fig. \ref{Fig3}b and \ref{Fig3}d) due to plasma defocusing. When $f_{probe} = f_{pump}$, the probe beam, which remains unaffected by the plasma effects at a delay of $\tau$ = 400 fs (Fig. \ref{Fig3}e and \ref{Fig3}g), presents a “central surge” at $\tau$ = -400 fs (Fig. \ref{Fig3}f and \ref{Fig3}h), while a light depletion or shadow is evident in the corona region (“corona shadow”).

A hypothesis for the different dynamics observed for the two focusing configurations is that the probe experiences a different nonlinear interaction in each case. The change in the focal position of the probe relative to plasma leads to a completely different interaction between pump and probe. This can be understood in terms of the effective refractive index variation experienced by the probe and the nonlinear effects induced by the pump, which are not the same in both situations. When the probe focuses before the pump ($f_{probe} < f_{pump}$), it suddenly experiences the plasma-induced decrease in the refractive index during the first stage of its divergent stage, leading to a different and more pronounce defocusing and creating a central shadow and a corona surge. By contrast, when both beams focus at the same position ($f_{probe} = f_{pump}$), the probe interacts with the plasma at its highest intensity and density (the focus). In this region, when the probe starts defocusing, it experiences the subsequent recovery or increase of the refractive index profile since the plasma is no longer present at that spatial position and it is even reinforced by the pump-induced Kerr nonlinearity. This combined effect can cause the medium to effectively act as a positive lens for certain spatial components, further enhancing the probe focusing and resulting in the observed "central surge”. 

Therefore, the relative focusing conditions between the pump and probe beams play a crucial role in the subsequent probe beam propagation. Also, in the two focusing scenarios considered here, the four intensity-modified identified spatial regions identified (shadow or surge at centre or corona) can be considered as an optical switch. In fact, in the literature, the PI-FROSt technique analyses intensity variations in the corona as a function of the pump-probe relative delay. Here, we extend this approach by analysing the behaviour of such a switch in the two different light collection regions (centre and corona) and the corresponding experimental and simulated PI-FROSt traces.

The experimental traces for the two focusing configurations are obtained recording the probe beam spectrum in the two light-collecting regions for different pump-probe delays. As previously noted, the selection of the collection region (centre or corona) is achieved by placing an iris or an anti-pinhole, respectively, before the collection optics (lens L2), which couples the light into an optical fibre connected to the spectrometer. The numerical traces are obtained by calculating the spectrum of the spatially integrated probe field in both spatial regions of light collection (centre and corona) at the propagation position under study ($z$ = 80 mm) and under both focusing conditions ($f_{probe} < f_{pump}$ and $f_{probe} = f_{pump}$). The trace of the probe beam in the centre (“centre trace”) is obtained by integrating a spatial area around the propagation axis, while the trace of the probe beam in the corona (“corona trace”) is obtained by integrating a spatial area at the periphery. The selected spatial region depends on the focusing and collection scenario. For the case in which $f_{probe} < f_{pump}$, a circular region of 0.22 mm radius is considered for the central spatial light collection, while in the corona case all signals beyond 0.5 mm from the optical axis are integrated. For the $f_{probe} = f_{pump}$ configuration, the central region is limited to a radius of 0.19 mm, while in the corona the signal is integrated from 0.8 mm outwards. These areas are chosen to maximize the signal contrast when the plasma is present and absent; nevertheless, the results are more sensitive to the spatial position of the integration region than to its size, especially in the corona case. 

\begin{figure}[htbp]
\centering\includegraphics[width=15cm]{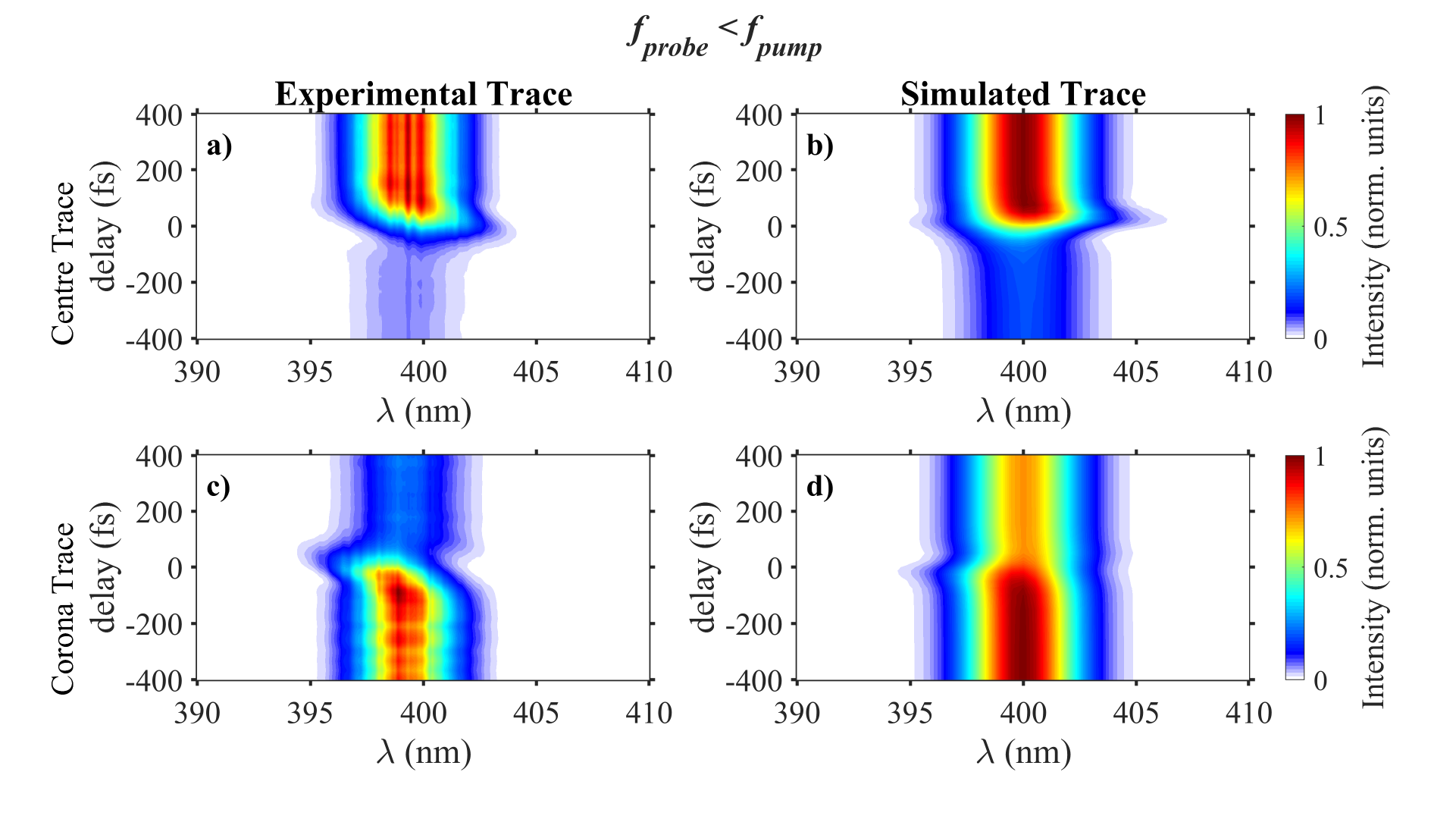}
\caption{Measured (a, c) and simulated (b, d) PI-FROSt traces collecting light from both the centre (a and b) and corona regions (c and f) using $f_{probe} < f_{pump}$. For negative delays, the probe pulse arrives after the pump, while for positive delays, it arrives before the pump.}
\label{Fig4}
\end{figure}

By applying these described criteria to collect light from both the central and corona regions, it is possible to obtain the PI-FROSt traces corresponding to the two focusing configurations studied. Figure 4 presents the PI-FROSt traces for the case where $f_{probe} < f_{pump}$: Experimental traces are shown in Fig. \ref{Fig4}a and \ref{Fig4}c, together with the corresponding simulations in Fig. \ref{Fig4}b and \ref{Fig4}d. The traces obtained from collecting light at the centre of the probe beam are shown in Fig. \ref{Fig4}a and \ref{Fig4}b; while Fig. \ref{Fig4}c and \ref{Fig4}d shows the traces corresponding to corona region light collection.

The traces obtained from the central region of the probe beam (Fig. \ref{Fig4}a and \ref{Fig4}b) exhibit the expected behaviour with a decrease in measured spectral intensity at negative pump-probe time delays, when plasma is present (i.e., at negative times, the probe beam arrives after the pump), due to plasma-induced defocusing. However, when light is collected from the corona region (Fig. \ref{Fig4}c and \ref{Fig4}d), the signal increases under the same conditions (i.e., the “corona surge” scenario), because the defocused light is redistributed toward the outer regions. Consequently, the choice of the spatial light collection zone (centre or corona) significantly impacts the trace shape, flipping it along the vertical direction, although it is not strictly symmetric and certain features differ. The simulated traces agree remarkably with the experimental results and also reflect the PI-FROSt trace direction flip when changing the spatial collection zone. Slight differences on the trace slope can be observed when comparing the experiments (Fig. \ref{Fig4}a and \ref{Fig4}c) to the simulations (Fig. \ref{Fig4}b and \ref{Fig4}d). The spectral cutoff is flatter in the simulated cases than in the experimental case, which can be attributed to the assumption of ideal Fourier-transform-limited pulses, while in the experiments the probe is slightly chirped by the optics ($\sim$ 1000 $fs^2$).

As shown in Fig. \ref{Fig4}, the signal contrast between the switched and non-switched situations is higher in the central region (Fig. \ref{Fig4}a and \ref{Fig4}b) than in the corona (Fig. \ref{Fig4}c and \ref{Fig4}d), where the contrast is reduced because the signal is integrated over a larger area where the unmodified pulse intensity remains significant, and it is also highly sensitive to the specific spatial region selected for the measurement.

\begin{figure}[htbp]
\centering\includegraphics[width=15cm]{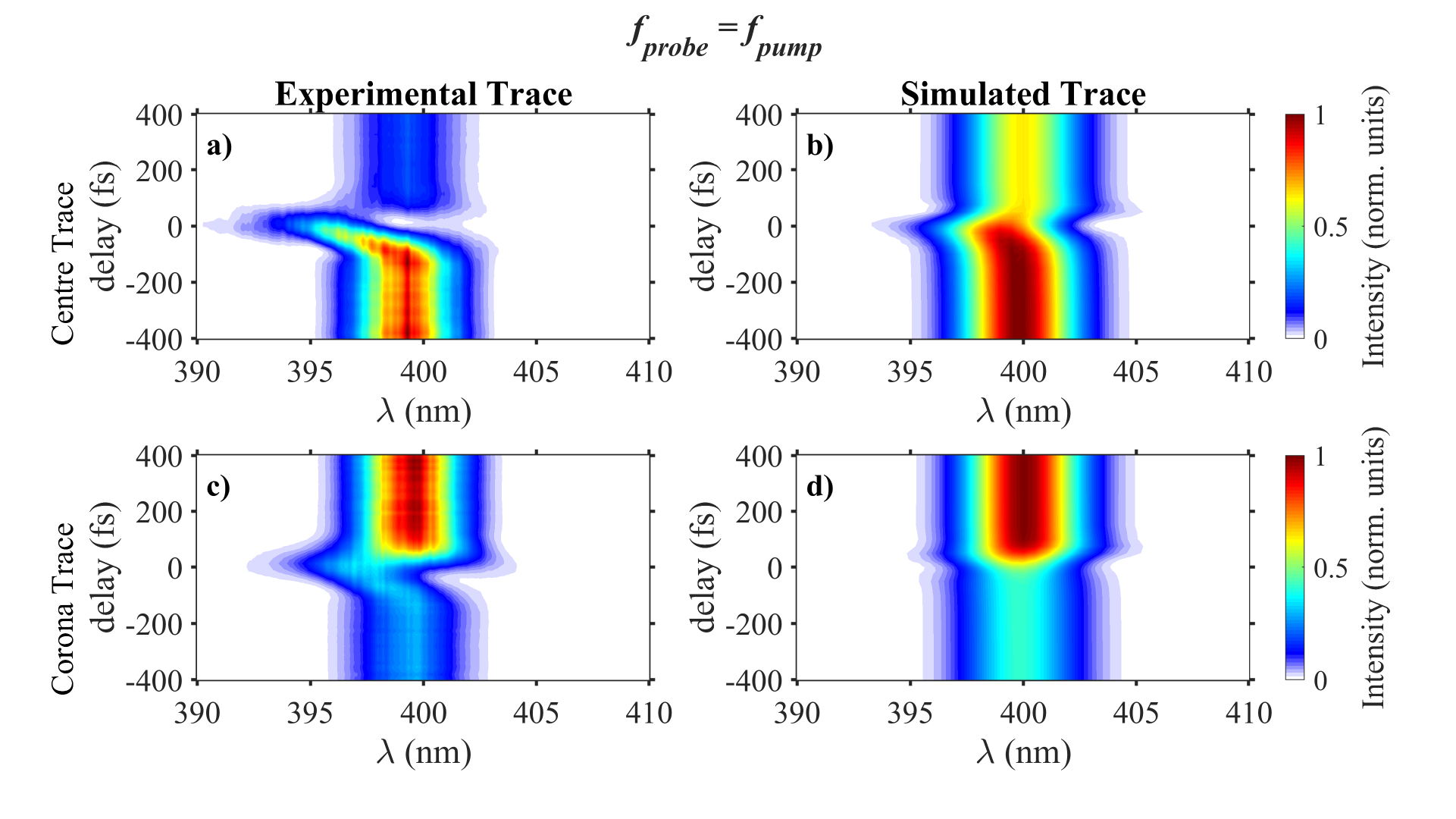}
\caption{Measured (a, c) and simulated (b, d) PI-FROSt traces collecting light from both the centre (a and b) and corona regions (c and d) using $f_{probe} = f_{pump}$. For negative delays, the probe pulse arrives after the pump, while for positive delays, it arrives before the pump.}
\label{Fig5}
\end{figure}

To highlight the impact and relevance of the focusing configuration regarding the PI-FROSt traces, we studied the same traces for the focal length configuration $f_{probe} = f_{pump}$ (Fig. \ref{Fig5}). Figure \ref{Fig5} shows both measured (Fig. \ref{Fig5}a and \ref{Fig5}c) and simulated (Fig. \ref{Fig5}b and \ref{Fig5}d) PI-FROSt traces. The most noticeable difference, compared to the case shown in Fig. \ref{Fig4}, is that the PI-FROSt trace directions are vertically flipped, which is consistent with the change from divergent to convergent-like behaviour observed in the probe dynamics, as mentioned earlier. 

This inversion corresponds to the propagation behaviour observed earlier in Figs. \ref{Fig2} and \ref{Fig3} when comparing the $f_{probe} < f_{pump}$ and the $f_{probe} = f_{pump}$ scenarios. The experimental PI-FROSt trace obtained from the central region of the probe beam (Fig. \ref{Fig5}a) shows a structure near zero delay, which is sharper than that observed in Fig. \ref{Fig4}a, with a more pronounced and opposite spectral shift (a clear blueshift in the $f_{probe} = f_{pump}$ scenario, Fig. \ref{Fig5}a, and a weak redshift in the $f_{probe} < f_{pump}$ scenario, Fig. \ref{Fig4}a). Also, the contrast between the switched and non-switched situations is lower than in Fig. \ref{Fig4}a. For the experimental PI-FROSt trace obtained from the corona region (Fig. \ref{Fig5}c), similar features can be observed: a flipped trace direction and a more pronounced structure with a sharper blueshift than in the case of $f_{probe} < f_{pump}$ (Fig. \ref{Fig4}c). The corresponding simulations (Fig. \ref{Fig5}b and \ref{Fig5}d) agree qualitatively with this behaviour.

The PI-FROSt traces also present a spectral shift of the probe pulse near zero delay due to the XPM effect induced by the pump pulse. For negative delays, the probe interacts mainly with the trailing edge of the pump ($dI/dt<0$). This produces a spectral shift of the probe spectrum towards shorter wavelengths (blueshift). For positive delays, the probe interacts with the leading edge of the pump pulse ($dI/dt>0$), resulting in a spectral shift towards longer wavelengths (redshift). The blueshift is only visible in the corona trace in Fig. \ref{Fig4}c and \ref{Fig4}d, since in the centre trace the signal is suppressed by the plasma-induced defocusing. In Fig. \ref{Fig5}, this effect becomes more significant because the probe interacts with a higher pump intensity (closer to pump focus).

In summary, the trace structure strongly depends on the focusing geometry of the pump and probe beams, as well as the light collection zone. In principle, obtaining different trace structures for the same probe pulse is not a limitation. Different interactions and, therefore, different gating conditions yield different traces; however, in theory, all of them encode the probe pulse information that can be retrieved with proper algorithms. Nevertheless, special attention must be given to potential measurement artefacts, which may mask or distort the encoded signal. Furthermore, it is essential to verify that the traces effectively correspond to the ideal definition: a gate function moving in time through a given signal. These points are addressed in the next subsection. 

\subsection{Gate function and its role in the trace}
In the PI-FROSt scheme, the interaction between the pump and probe pulses is governed by a temporal gate arising from the nonlinear response of the medium induced by the pump pulse, which switches the test pulse under analysis (probe pulse). In particular, the gate function results from the combined action of cross-phase modulation (XPM) and the plasma-induced change in the refractive index, along with its impact on the probe beam propagation. Under the usual assumptions of gating-based retrieval techniques, the detected signal can be expressed as the product of the field to be characterised, $E(t)$, and a gate function $g(t)$ that encodes the temporal response of the medium: 
\begin{equation}
S(t,\tau) \approx E(t)\cdot g(t-\tau)
\label{Eq6}
\end{equation}

If this condition holds, the pulse can be reconstructed using the standard retrieval methods mentioned previously. 
Therefore, in order to work properly, the PI-FROSt setup should deliver a trace following the condition described by Eq. (\ref{Eq6}). To assess whether this requirement is fulfilled in our case, we analyse the effective gate extracted from the simulations. Since the numerical model provides access to both the original pulse and the gated signal at each delay, we can directly retrieve $g(t)$ by calculating the ratio between the complex fields with and without the pump interaction, integrated over the selected spatial region (centre or corona). If the underlying assumption is valid, the extracted gate should exhibit the same temporal shape for all delays, differing only by a temporal shift given by the delay scan. Conversely, if this behaviour is not observed, the assumption breaks down and the reconstruction procedure is no longer exact, and the retrieved pulse is only approximated. 

\begin{figure}[htbp]
\centering\includegraphics[width=15cm]{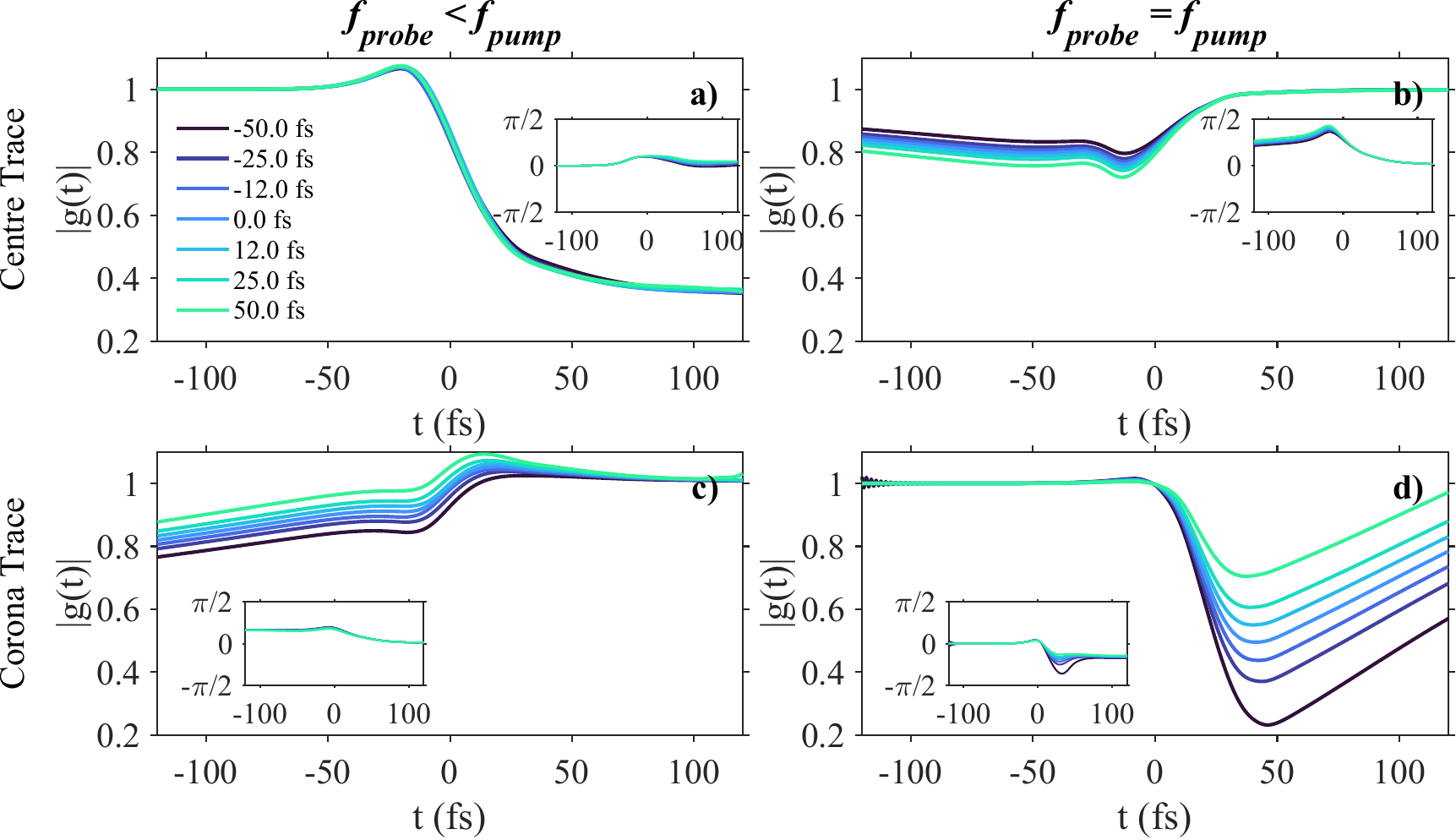}
\caption{Magnitude of the retrieved gating functions for the $f_{probe} < f_{pump}$, (a and c) and $f_{probe} = f_{pump}$ (b and d) configurations. Panels a and b: light collection measurement in the central region. Panels c and d: light collection in the corona region. Only delays with temporal overlap between pump and probe are displayed. Insets show the gate phases in each case.}
\label{Fig6}
\end{figure}

Figure \ref{Fig6}a and \ref{Fig6}c shows the magnitude of the retrieved gates for the $f_{probe} < f_{pump}$, while the corresponding phases are displayed in the inset. The gate related to the light collection at the centre (Fig. \ref{Fig6}a) is obtained by dividing the probe electric field at each delay by the probe field measured in the absence of plasma (both integrated over the same spatial region around the optical axis where the centre PI-FROSt trace has been obtained). As a result, the absolute value of the gate is 1 when the probe does not interact with the plasma and drops below 0.4 when the signal is depleted due to the plasma-induced defocusing in the central region. A modest increase of the gate above 1 is observed at delays close to $-20\,\mathrm{fs}$ indicating the presence of residual XPM effects. This temporal profile matches the expected nonlinear phase shift induced by XPM and plasma dynamics. Conversely, the gates related to the corona collection scenario (Fig. \ref{Fig6}c) are calculated by dividing the probe electric field at each delay to the probe field at a fixed delay where it interacts with the plasma, both integrated over the same spatial region in the periphery. In this way, the gate takes an absolute value of 1 when the probe is defocused by the plasma and decreases when the probe does not interact with it, reflecting the reduction of the corona signal in the absence of defocusing.

The magnitude of the retrieved gates functions for the case $f_{probe} = f_{pump}$ is shown in Fig. \ref{Fig6}b and \ref{Fig6}d. Consistent with the traces, the gates are reversed with respect to the previous case. The gate from the centre collection scenario (Fig. \ref{Fig6}b) is obtained by dividing the probe electric field at each delay by the probe field interacting with the plasma at a fixed delay, yielding an absolute value of 1 when the plasma is present and decreasing when it is absent. This behaviour reflects the signal enhancement in the central region due to nonlinear focusing. The gates obtained from the corona collection scenario (Fig. \ref{Fig6}d) are calculated by dividing the probe electric field at each delay to the probe field propagating in the absence of plasma. In this case, the gate takes a value of 1 when the plasma is not present and decreases when the probe interacts with the plasma, reflecting the reduction of the corona signal due to focusing. 

There is a clear difference between gates obtained from the centre and corona collection regions: while in the central collection configuration all the retrieved gates present the same temporal shape (up to the time shift introduced by the delay, which is subtracted for a better comparison), in the corona case the gates exhibit noticeable variations depending on the delay and the specific spatial region selected for the measurement. The results suggest that, in the central collection region, the effect is essentially the same in both near- and far- field regimes. In contrast, in the corona region the effect only becomes apparent in the far field, indicating a dependence on diffraction and pulse propagation. As a consequence, the retrieved gate depends on the portion of the probe pulse interacting with the plasma, which changes with the pump-probe delay. The gate depends not only on whether the signal is collected from the centre or the corona region, but also on the specific spatial subregion chosen within each of them, which affects both the retrieved traces and the corresponding gates.

The information shown in Fig. \ref{Fig6} suggests that the centre gate obtained for the case $f_{probe} < f_{pump}$ provides the most stable and robust gate for each delay, both in terms of its amplitude and the associated phase (see the inset of Fig. \ref{Fig6}a). In some cases, the phase remains small, suggesting that it may not be necessary to take it into account in the pulse retrieval process. This interpretation is consistent with previous studies, in which the switching function was found to behave predominantly as an intensity filter, not because the gate is strictly real, but because the phase contribution does not appear to play a relevant role in the retrieval process \cite{Leblanc2019}.

Therefore, from the four possible traces, it seems that in the $f_{probe} < f_{pump}$ case, central light collection corresponds most closely to the condition given by Eq. (\ref{Eq6}), while the rest of the cases add relevant distortions to the trace structure that should be addressed by an appropriate retrieval algorithm for proper pulse characterisation. 

Please note that this study considers a collinear pump and probe approach, while in the literature PI-FROSt is typically performed in non-collinear setups: thus, the conclusions are derived within the framework of the cases considered here.
However, the main message from the present results is that the PI-FROSt trace and its quality strongly depend on the pump and probe focusing conditions and on the light collection scheme. Therefore, for any given setup configuration, the impact of the focusing geometry on the trace and gate properties should be carefully considered.

\subsection{Dependence on the probe wavelength}
\begin{figure}[htbp]
\centering\includegraphics[width=15cm]{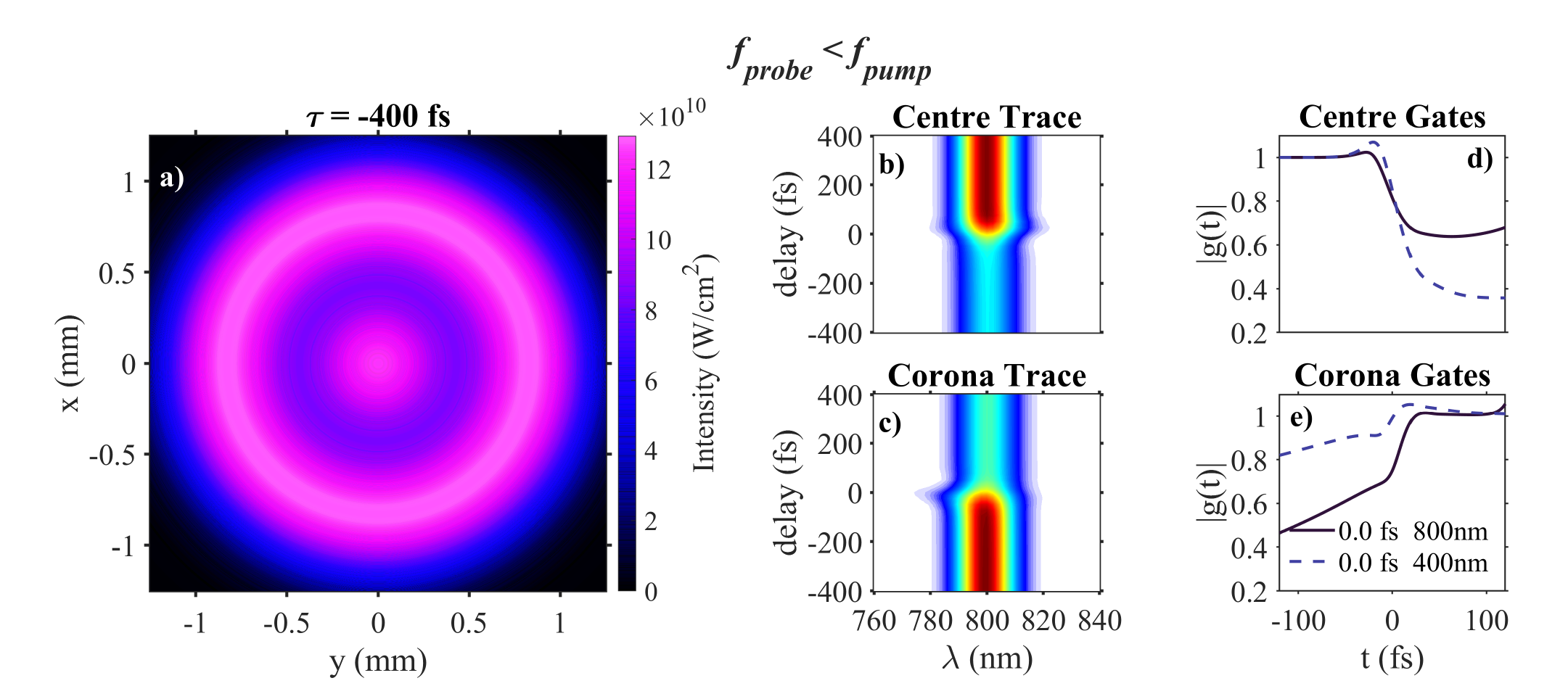}
\caption{Simulated probe transverse intensity profile at 800 nm (a) after interacting with the same plasma for the case $f_{probe} < f_{pump}$. Simulated PI-FROSt traces collecting light both in the central (b) and corona regions (c). For negative delays, the probe pulse arrives after the pump, while for positive delays, it arrives before the pump. Retrieved gating functions for centre (d) and corona (e) traces at zero delay. The corresponding gates obtained for a probe wavelength of 400 nm are shown as dashed lines.}
\label{Fig7}
\end{figure}

Finally, we have numerically addressed the dependence of the plasma defocusing with the probe pulse wavelength. We have considered a probe pulse with the same parameters as in the previous sections, varying only its central wavelength to 800 nm, which is the same as the pump wavelength. This allows to compare the plasma’s effect on the probe propagation dynamics at different probe wavelengths, while keeping the plasma conditions unchanged. Again, we have considered perpendicular polarisation relative to the pump beam to minimise XPM and interferences. We have simulated the case where $f_{probe} < f_{pump}$. Figure \ref{Fig7}a shows the simulated transverse intensity profile of the probe beam centred at 800 nm (i.e., time integrated) at a distance of $z$ = 80 mm for a delay of $\tau$ = -400 fs (probe after pump). The simulated PI-FROSt traces are obtained by calculating the spectrum of the probe field spatially integrated over the central (Fig. \ref{Fig7}b) and corona (Fig. \ref{Fig7}c) regions, for the different pump-probe time delays. The areas in which spatial integration is performed are adjusted to obtain the maximum possible contrast. The analysis of the probe pulse evolution reveals that the probe experiences significantly more defocusing by the plasma at 800 nm than at 400 nm, obtaining a better contrast in the corona trace (compare with Figs. \ref{Fig3} and \ref{Fig4}). 

This behaviour is expected, since the plasma critical density ($\rho_c$) scales with $\lambda^{-2}$. As a result, the critical density for the same plasma is lower at 800 nm than at 400 nm. Considering that the plasma-induced refractive index change is inversely proportional to this critical density ($\Delta n \propto -\rho_C^{-1}$) \cite{COUAIRON2007} then, for a given plasma density, the 800 nm pulse experiences a larger refractive index change, which directly results in a more pronounced defocusing. In addition, the features of the beam at the focal point also differ, with the 800 nm beam size twice as large as that of the 400 nm beam for the same plasma. This explains why some light is still visible at the beam centre in the transverse intensity profile (Fig. \ref{Fig7}a), which leads to a lower contrast PI-FROSt trace (Fig. \ref{Fig7}b) compared to that obtained with a probe wavelength of 400 nm. An analogous argument applies to the corona PI-FROSt trace (Fig. \ref{Fig7}c): the stronger probe diffraction at 800 nm results in a higher signal contrast between the presence and absence of plasma conditions. The change in the probe wavelength has a direct impact on the gate shape. Fig. \ref{Fig7}d and \ref{Fig7}e show the retrieved gating functions for the traces obtained in the centre and corona collection scenarios, respectively, at zero delay. The corresponding gates obtained with a 400 nm probe are shown as dashed lines for direct comparison. A change in the probe wavelength results in a different gating function, as the interaction between the probe and the plasma is modified, reflecting the wavelength dependence of the depletion and diffraction processes that determine the effective temporal gating. This result indicates that the gating response is not universal but depends not only on the spatial region of the beam but also on the probe wavelength. Consequently, reconstruction algorithms should consider this dependence, especially when dealing with a beam including fundamental and harmonic wavelength components, since assuming a fixed or wavelength-independent gate may introduce inaccuracies in the retrieved pulse characteristics.

\section{Conclusions}
In this work, we have studied the Plasma-Induced Frequency-Resolved Optical Switching (PI-FROSt) technique from a fundamental perspective, combining a (2+1)D numerical model with experimental measurements. Our results demonstrate that the performance and characteristics of the PI-FROSt traces are highly sensitive to the relative focusing geometry of the pump and probe pulses, as well as the spatial region selected for signal collection. 

We identified two different scenarios related to the probe behaviour: a plasma-induced defocusing effect when probe focuses before the plasma, and a focusing (positive lens) effect when both are focused at the same position. These interactions result in four intensity-modified spatial regions, characterised by the shadowing or surging of the light signal in both the centre and peripheral (corona) regions. We have shown that the selected signal collection zone can flip the trace direction and significantly impact the interpretation of the measured experimental data.

The (2+1)D numerical model reproduced these findings, showing qualitative agreement with experimental data and proving its reliability for exploring nonlinear dynamics. Furthermore, we have used numerical simulations to study the temporal gate function $g(t)$, observing that the most defined and stable gate $g(t)$ is obtained when the probe is focused before the plasma and the signal is collected at the central region of the beam. In addition, the temporal gate function $g(t)$ may have a phase and depends on the probe wavelength, as the propagation, focusing and the effect induced by the plasma vary strongly with the frequency. 

PI-FROSt is a robust technique for broadband pulse characterisation, with a wide range of demonstrated applications. This work highlights that precise spatial and temporal optimisation of the focusing geometry and the collection scheme is essential to avoid measurement artefacts and ensure accurate pulse reconstruction.

\section*{Acknowledgments}
Authors thank support from the Department of Education of the Junta de Castilla y León and FEDER Funds UE (SA108P24, Escalera de Excelencia CLU-2023-1-02) and Ministerio de Ciencia, Innovación y Universidades (MICIU/AEI/10.13039/501100011033, project PID2023-149836NB-I00). I.J.S. and M.G. performed the experiments. A.C. developed the model and performed simulations. All authors contributed to the discussion of the results and writing.

\bibliographystyle{unsrt}
\bibliography{bibPIFROST}

\end{document}